\documentclass{aa}  
\usepackage{graphicx}
\usepackage{gensymb}
\usepackage{amsmath}
\usepackage{txfonts}
\begin{document}

   \title{Amplification of active region equatorial dipole}
   
   \author{Ismo Tähtinen
                 \inst{1}
         % \and
          }
   \institute{Space Physics and Astronomy Research Unit, University of Oulu,
               POB 8000, FI-90014, Oulu, Finland\\
              \email{ismo.tahtinen@oulu.fi}
    %     \and
     %        University of Alexandria, Department of Geography, ...\\
      %       \email{c.ptolemy@hipparch.uheaven.space}
       %      \thanks{The university of heaven temporarily does not
        %             accept e-mails}
             }

   %\date{Received September 15, 1996; accepted March 16, 1997}

% \abstract{}{}{}{}{} 
% 5 {} token are mandatory
 
  \abstract
   {The effect of active region parameters on the evolution of solar axial dipole is well understood.
   However, their effect on the solar equatorial dipole has not been studied in detail.
   Understanding the development of the equatorial dipole is important as it drives the interplanetary magnetic field on intracyclic timescales ($\sim $1yr).
}
   {We study how the latitude, tilt angle, and the polarity separation affect the evolution of solar equatorial dipole.
   }
   {We use dipole flux transport (DFT), a matrix implementation of surface flux transport (SFT) model, to simulate the dipole evolution of synthetic bipolar magnetic regions (BMRs) with a wide range of tilt angles and latitudes.
   We also study the evolution of active regions in HMI SHARP database and their associated BMRs.
   We quantify the dipole evolution by means of the maximum and mean amplification and the growth and decay times of the equatorial dipole.
   }
   {Tilt angle controls the amplification of the equatorial dipole, with larger tilt angles leading to stronger equatorial dipole.
   Latitude controls the timescale at which the equatorial dipole grows and decays.
   Low-latitude regions produce the longest living equatorial dipole.
   %We find a considerable asymmetry between regular and anti-Joy regions.
   We find that only 31\% of anti-Joy regions experience amplification of equatorial dipole, compared with 96\% of regularly tilted active regions.
   }
   {Latitudes and tilt angles of active regions can have considerable effect on the solar equatorial dipole.
   Systematic changes in their properties could increase/decrease the strength of the equatorial dipole relative to the solar activity, which could lead to similar relative change in geomagnetic activity.
   We suggest that large low-latitude coronal holes and relatively high geomagnetic activity in the declining phase of solar cycle 20, could be related to emergence of unusually low-latitude active regions that produced strong and persistent equatorial dipole. 
   }


   \maketitle
%
%-------------------------------------------------------------------

\section{Introduction}\label{sec:Intro}
The global solar magnetic field is the large-scale magnetic structure of the Sun that is rooted in photospheric flux and shapes the structure of the heliosphere.
It is often modeled as solar dipole and it is primarily shaped by emergence of strong magnetic flux bundles, active regions, that arise from the convection zone and disperse across the photosphere with flux transport processes.
Because active regions are the source of the long-lived photospheric magnetic fields, the evolution of the global solar magnetic field is intrinsically tied to their properties.
While the connection between the active regions and build-up of the solar axial dipole is rather well understood, the relationship between the active regions and the evolution of the solar equatorial dipole has barely been studied.

Axial dipole has enjoyed a widespread interest due to its connection with Babcock-Leighton mechanism that relates the strength of the polar fields at the solar minimum to the amplitude of subsequent solar cycle.
This connection can be used to forecast the amplitude of the subsequent solar cycle and also opens up a possibility that the systematic changes in active region properties might regulate the solar dynamo \citep{Schatten1978,Svalgaard2005,Cliver2011,Upton2014,Cameron2016,Ijima2017,Jiang2018,Whitbread2018,Upton2018,Bhowmik2018,Yeates2020,Petrovay2020,Jiang2023,Yeates2025}.
While the equatorial dipole component is much less studied, it plays an important role in shaping the interplanetary conditions mid-term timescales from a few solar rotations to a few years \citep{WangSheeley2000a,WangSheeley2000b,WangSheeley2003}.
For example, there is substantial evidence that the rapid strengthening of the heliospheric magnetic field, and the resulting open flux peak in late 2014, were primarily driven by the strong equatorial dipole caused by the persistent emergence of active regions within a southern-hemisphere longitude band over several solar rotations \citep{WangSheeley2015,Tahtinen2026a,Yoshida2026}.

In an ideal case, active region appears as a bipolar magnetic region (BMR) that consist of two magnetic patches of equal flux and opposite polarity that correspond to the opposite ends of the flux bundle threading the photosphere.
BMRs are typically characterized by five parameters: total magnetic flux, latitude, longitude, tilt angle with respect to the equator, and the separation distance between poles.
The evolution and effect of individual BMRs on the net solar dipole can be simulated with surface flux transport  \citep[SFT, see, e.g.,][]{Yeates2023} models.
Pioneering work on the subject was done by \citet{WangSheeley1991} who laid out the basic principles of how the properties of individual regions connect to the evolution of the net dipole.

In the classic SFT model, the photospheric magnetic field evolves under the supergranular diffusion and two large-scale flows, the differential rotation and the poleward driven meridional flow.
The evolution of the BMR magnetic flux in SFT simulations can be quantified with a solar dipole, which corresponds to first multipole in the spherical harmonic expansion of photospheric field. 
As the classic SFT model is linear, the net solar dipole can be obtained by vectorially summing individual BMR dipoles \citep{WangSheeley2000b,Tahtinen2026a}.
The net axial component is even more straightforwardly just the regular (scalar) sum of the individual axial components.

For the axial dipole, the most consequential parameters are the latitude and tilt angle of a BMR.
Tilt angle is especially important as it breaks the symmetry between the BMRs poles by providing latitudinal separation between the opposite polarities \citep{WangSheeley1991}.
Due to the latitudinal separation, more of the lower latitude (typically leading polarity) flux will diffuse equatorward and depending on the latitude may cross the equator, causing an anti-symmetric distribution of flux across the hemispheres.
The magnetic flux that crosses the equator will then be carried away by the oppositely directed meridional flow, which will cause the axial dipole to grow and after few years reach an asymptotic steady state in which the remaining opposite polarity flux resides on poles.
For BMRs, there is a particularly simple relation: the asymptotic strength of the axial dipole can be described in terms of amplification factor of the initial axial dipole, which is a Gaussian function of emergence latitude only \citep{Jiang2014,Petrovay2020}.

While the axial dipole is amplified through a combined process of diffusion and meridional flow, the equatorial dipole is amplified by the differential rotation \citep{WangSheeley2000b,WangSheeley2002}.
Similarly to axial dipole, a nonzero tilt angle is necessary for there to exist latitudinal separation between the polarities that the latitudinal shear due to differential rotation can tear apart in longitude.
The equatorial dipole of BMR is short-lived and eventually decays as there is no mechanism that could sustain the longitudinal separation between opposite polarities \citep{WangSheeley2003}.
Although the equatorial dipole is rather short-lived, it typically experiences quite strong amplification during its lifetime.
\citet{WangSheeley1991} find that the equatorial dipole component reaches its maximum strength in about half years and then rapidly decays on a timescale of about one year.
While the longitude at which BMR emerges does not affect the evolution of single BMR, the best-known net effect of multiple BMRs on the net equatorial dipole is the effect of their longitude distribution.
Since the direction of the equatorial dipole component of BMR depends on its longitude, the dipole components of multiple BMRs can interfere both constructively and destructively depending on their longitudinal distribution.
For this reason, the equatorial dipole of many uniformly distributed BMRs is small compared to many BMRs emerging at nearby longitudes of one hemisphere.

While it is known that a non-zero tilt angle is necessary for the equatorial dipole of BMR to grow, how exactly the BMR parameters affect the equatorial dipole evolution has not been studied in detail.
In this paper we study how the tilt angle, latitude and polarity separation affect the evolution of BMR equatorial dipole with dipole flux transport \citep[DFT, ][]{Tahtinen2026b}, which is a extremely efficient matrix implementation of the SFT model.
We also study the equatorial dipole evolution of active regions in HMI SHARP dataset and how the BMR approximation affects their evolution.
In Sect.~\ref{sec2:Data} we present the data.
Section~\ref{sec4:DFT} presents the DFT model and the dipole vector representation that we use to simulate and quantify the evolution of BMRs.
In Sect.~\ref{sec5:CharacteristicEvolution}, we present four characteristic examples of dipole evolution.
In Sect.~\ref{sec6:EffectOfParameters} we analyze the effect of tilt angle, latitude and polarity separation on the evolution of equatorial dipole by simulating the evolution of thousands of synthetic BMRs.
In Sect.~\ref{sec7:SHARPS} we analyze the evolution of active regions in HMI SHARP dataset and as well as corresponding BMR's.
We discuss our results in Sect.~\ref{sec:Discussion} and give our conclusions in Sect.~\ref{sec:Conclusions}.

\section{Data}\label{sec2:Data}

\subsection{HMI SHARPs and bipolar magnetic regions}
We use an active-region database derived from Spaceweather HMI Active Region Patch data \citep[SHARP;][]{Bobra2014}, using an open-source code\footnote{\url{https://github.com/antyeates1983/sharps-bmrs}} described in \citet{Yeates2020}.
This algorithm transforms HMI SHARPs onto a uniform sine-latitude and longitude grid and removes active regions with a high flux imbalance.
Active regions that are recognized as remnants of regions observed in the previous rotation are removed.
The data consist of 1108 active regions over solar cycle~24 (CR2096--CR2224) and 858 active regions for the ongoing solar cycle~25 (CR2225--CR2301).

We also use the same open-source code of \citet{Yeates2020} to transform HMI SHARPs into BMRs and to create synthetic BMRs from arbitrary BMR parameters.
The BMR algorithm includes a scaling parameter $a$ that controls the size of the BMR relative to the separation of the BMR poles.
\citet{Yeates2020} set $a=0.56$ in order to match the axial dipole moments of the derived BMRs with those of the original HMI SHARPs.
We adopt the same value of $a$ as we find that it also works for the equatorial dipole components of active regions, with the slope of the linear relation between the initial equatorial components of the HMI SHARPs and the corresponding BMRs being 1.004.

\subsection{BMR parameters}
BMRs are typically characterized by five parameters total flux $\Phi$, central latitude $\lambda$, central longitude $\phi$, tilt angle $\gamma$, and polarity separation $\rho$.
Here we focus on $\lambda$, $\gamma$, and $\rho$, which, following \citet{Yeates2020},  are defined as follows.
The BMR centroid in the sine-latitude coordinate is 
\begin{align}
    s_0 = \frac{1}{2}\left(s_++s_-\right),
\end{align}
where $s_+$ and $s_-$ denote the sine-latitudes $\sin{\lambda_+}$ and $\sin{\lambda_-}$ of positive and negative polarities, respectively.
The polarity separation, $\rho$, is the heliographic angle between the two polarities and is given by
\begin{align}
    \rho =
    \arccos\left[
        s_+s_-
        + \sqrt{1-s_+^2}\sqrt{1-s_-^2}
        \cos\left(\phi_+ - \phi_-\right)
    \right],
\end{align}
where $\phi_+$ and $\phi_-$ denote the longitudinal locations of positive and negative polarities.

We modify the tilt-angle definition of \citet{Yeates2020} by introducing the hemispheric and cycle sign factors, $\mathrm{sign}(s_0)$ and $\eta$, to obtain a cycle- and hemisphere-agnostic tilt angle.
Latitudinal and longitudinal separations, $\Delta\lambda$ and $\Delta\phi$, are then 
\begin{align}
    \Delta \lambda &= \eta(\lambda_+-\lambda_-) \\
    \Delta \phi &= \mathrm{sign}(s_0)\,\eta\sqrt{1-s_0^2}\,(\phi_--\phi_+),
\end{align}
where
\begin{align}
\eta =
\begin{cases}
    1, & \text{for an odd cycle}, \\
    -1, & \text{for an even cycle}.
\end{cases}
\end{align}
The factors $\mathrm{sign}(s_0)$ and $\eta$ account for the alternation of BMR polarities between hemispheres and between odd and even cycles, and the factor $\sqrt{1-s_0^2}$ accounts for the decrease in the physical longitudinal distance with latitude.
The tilt angle, $\gamma$, with respect to the equator is then given by
\begin{gather}\label{eq:TiltAngle}
\gamma = \operatorname{atan2}(\Delta \lambda,\Delta \phi).
\end{gather}

\begin{figure} 
\resizebox{\hsize}{!}{\includegraphics{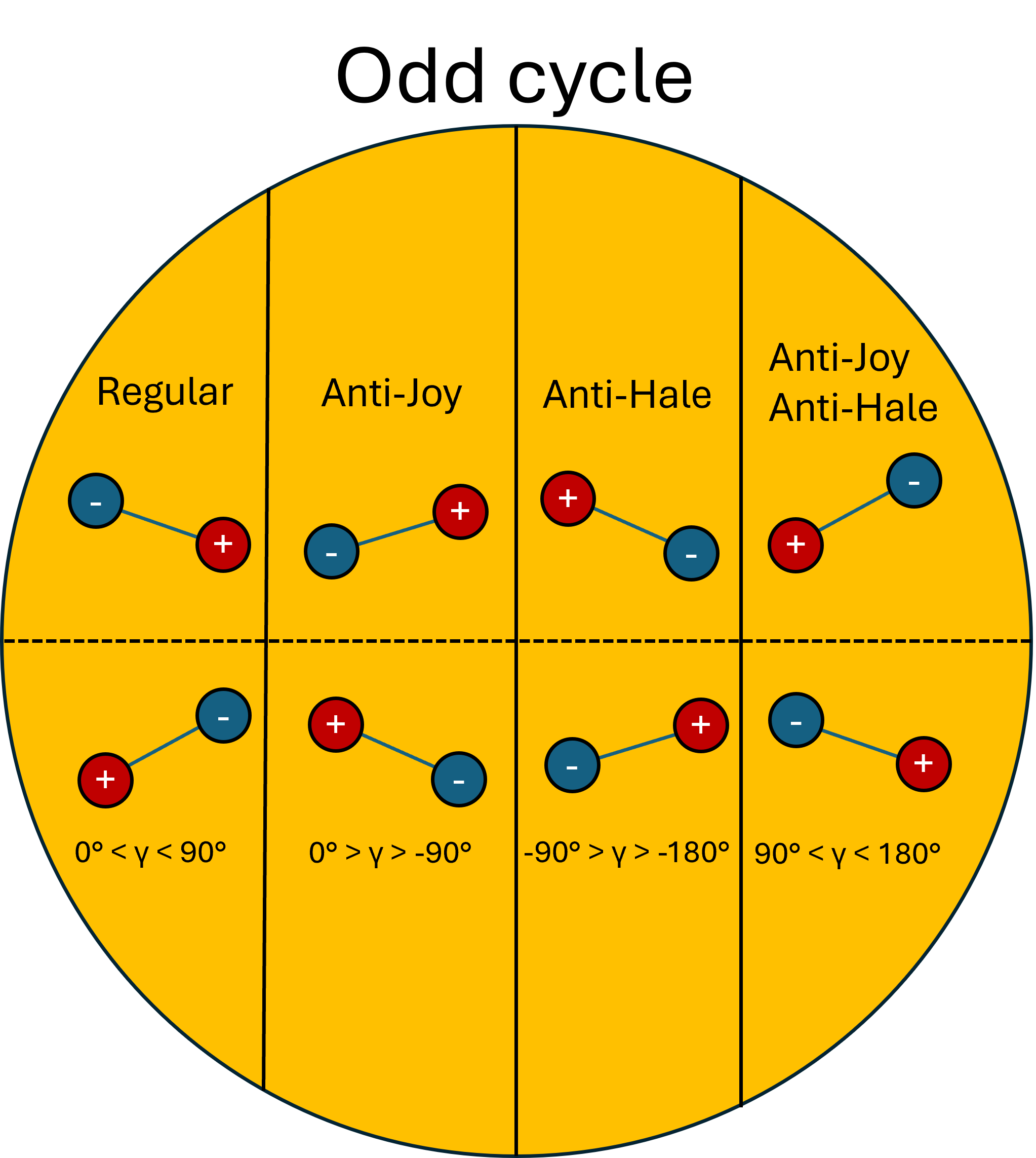}}
\caption{Visualization of the tilt angles for the odd solar cycle. Red circles correspond to positive and blue circles to negative polarities. Dashed line marks the solar equator.
}\label{fig:TiltVisualization}
\end{figure}

According to this definition, active regions obeying both Joy's and Hale's laws always have a positive tilt angle between 0\degree{} and 90\degree{}.
Similarly, anti-Joy regions have a negative tilt angle between -90\degree{} and 0\degree{}, while anti-Hale regions have a negative tilt angle between -180\degree{} and -90\degree{}.
The last quadrant, which is both anti-Joy and anti-Hale, has positive tilt angles between 90\degree{} and 180\degree{}.
The tilt-angle definition is illustrated for the odd cycle in Fig.~\ref{fig:TiltVisualization}.

\section{Dipole simulations}\label{sec4:DFT}
\subsection{Dipole flux transport}
We use dipole flux transport \citep[DFT\footnote{\url{https://github.com/itahtine/DipoleFluxTransport}},][]{Tahtinen2026b} model to simulate the evolution of BMRs and active region SHARPs.
DFT is a matrix formulation of the linear, time-independent surface flux transport (SFT) model described in Appendix~\ref{appendix}.
In DFT, numerical integration is replaced with a matrix multiplication
\begin{equation}\label{eq:VSProduct}
\mathbf{D}(t) = \mathbf{P}(t)\mathbf{B}_0,
\end{equation}
where $\mathbf{P}(t)$ is the propagator matrix that maps the initial magnetic field map $\mathbf{B}_0$ (e.g., active region) to the dipole vector at later time $t$.
We have constructed the set of such propagator matrices with daily resolution, which is sufficient to capture the evolution of the equatorial dipole in detail.

The propagator matrices are constructed by simulating the evolution of unit elements of a synoptic map with the SFT model and then representing this evolution with a dipole vector obtained using the recent vector sum method \citep{Tahtinen2024,Tahtinen2026a}.
DFT retains only the information needed to describe the dipole evolution, while preserving the underlying SFT dynamics at the temporal resolution of the propagator matrices.
Because DFT is much faster than the full SFT model, it enables a large number of simulations with negligible computational cost while providing identical results for the dipole evolution.

\subsection{Open dipole flux vector}\label{sec3:DipoleVector}
The dipole vector $\textbf{D}$ returned by the DFT model corresponds to the vector produced by the vector sum method of \citet{Tahtinen2024,Tahtinen2026a}.
It shares orientation with traditional solar dipole, but has different magnitude and units of magnetic flux instead of flux density.
Magnitude \textbf{|\textbf{D}|}, which we call open dipole flux (ODF), is related to the total solar dipole moment by a factor of $\frac{4\pi R^2_\odot{}}{3}$.
Importantly, ODF closely matches the open solar flux (OSF) from the potential field source surface (PFSS) model with a standard source surface radius of $R_{ss} = 2.5R_\odot{}$, while also equaling the total photospheric magnetic flux aligned with the magnetic dipole axis.
ODF thus directly relates the PFSS OSF to the magnetic flux distribution at the photosphere.

\subsection{Studied quantities}
In this paper, we study the effect of BMR parameters on the subsequent evolution of BMR magnetic flux.
We focus on the evolution of their equatorial dipole, which we quantify using the equatorial component of the ODF vector
\begin{align}
    D_{\rm eq}=|\textbf{D}|\cos{\lambda},
\end{align}
where $\lambda$ is the latitude of the dipole.
We mostly focus on the amplification factor, $A$, of the equatorial dipole, which we define by normalizing the equatorial component with the initial total dipole strength of BMR, $D_0=|\textbf{D}(t=0)|$,
\begin{align}
A(t)=D_{\rm eq}(t)/D_0.
\end{align}
We normalize by the initial total dipole strength rather than the initial equatorial component because $D_{\rm eq}(t=0)$ vanishes for highly tilted BMRs.

For characterizing the maximum strength of the equatorial dipole, we define the maximum amplification factor, $A_{\rm max}$, as
\begin{align}
A_{\rm max}= \max(D_{\rm eq}(t))/D_0.
\end{align}
For characterizing the overall strength of the equatorial dipole, we define mean amplification, $ \langle A \rangle $, as
\begin{align}
    \langle A \rangle = \frac{1}{T+1}\sum_{t=0}^{T}D_{\rm eq}(t)/D_0.
\end{align}
Here $T$ corresponds to timescale over which we measure the amplification of the equatorial dipole. 
We take $T$ to be 600 days which equals 22 Carrington rotations, and corresponds to maximum decay time, $t_{\rm decay}$, in our BMR simulations.
Decay time is defined as the time it takes for equatorial dipole strength to permanently decline below the initial dipole $D_0$. 
We also define growth time, $t_{\rm growth}$, as the time of the peak equatorial dipole strength. 

\section{Characteristic examples of active region dipole evolution}\label{sec5:CharacteristicEvolution}
\subsection{HMI SHARPs}
\begin{figure*} 
\resizebox{\hsize}{!}{\includegraphics{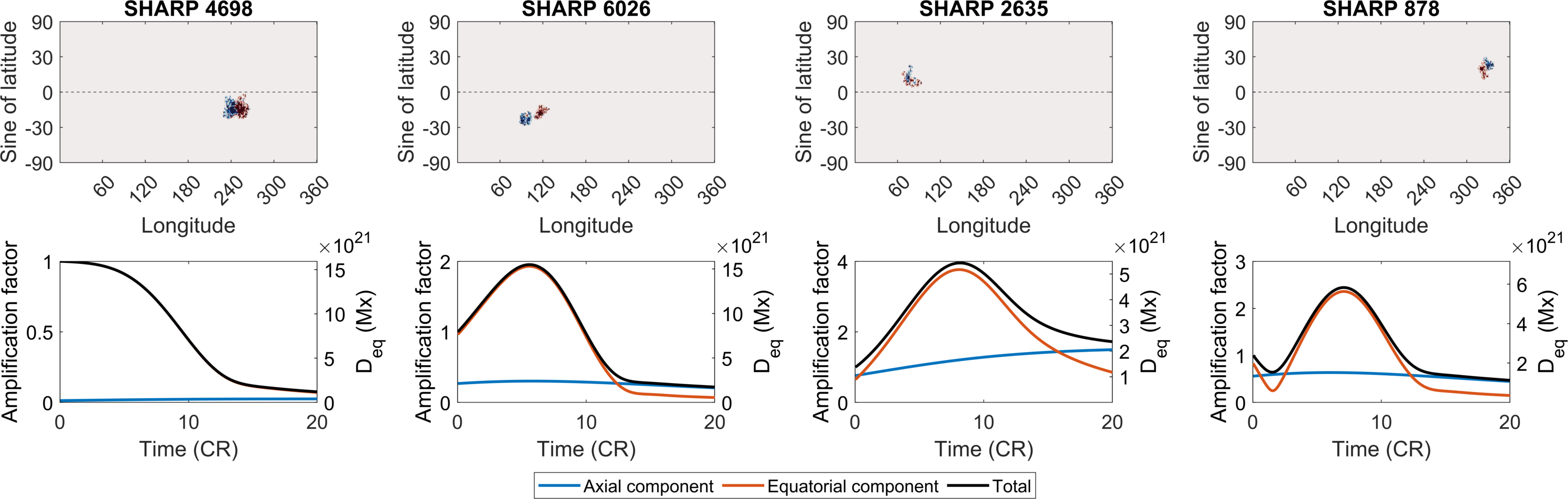}}
\caption{Evolution of active region dipole components. Top row show shows the active region and the bottom row the evolution of dipole components. The black line shows the evolution of the total dipole strength while the blue and orange lines show the evolution of axial and equatorial components.
}\label{fig:ARExamples}
\end{figure*}

Figure~\ref{fig:ARExamples} shows the evolution of the dipole components for four HMI SHARPs selected to illustrate different patterns of equatorial dipole evolution.

The active region in the first column of Fig.~\ref{fig:ARExamples} is HMI SHARP 4698, which corresponds to NOAA AR 12192, the strongest active region in 24 years, that emerged in October 2014.
The equatorial component dominates the evolution of this active region because of its nearly zero tilt angle.
The dipole begins to decay immediately after the emergence of the active region and undergoes no amplification.

In contrast, HMI SHARP 6026 in the second column of Fig.~\ref{fig:ARExamples} provides an example of moderate amplification. Its maximum dipole strength is about twice its initial strength. As a result, its maximum dipole strength reaches $1.5\times10^{22}$~Mx, which is about the same as the maximum strength of SHARP 4698, although the total magnetic flux of the SHARP 6026 is only 28\% of that of SHARP 4698.

HMI SHARP 2635, shown in the third column of Fig.~\ref{fig:ARExamples}, experiences even stronger amplification. 
Its maximum dipole strength is almost four times its initial strength.
Because of its large tilt angle, the equatorial component of this active region is initially smaller than its axial component.

Finally, HMI SHARP 878, shown in the fourth column of Fig.~\ref{fig:ARExamples}, displays a peculiar evolution where the dipole strength initially decreases, but is then amplified to more than twice of the initial value.

\subsection{Bipolar magnetic regions}
\begin{figure*} 
\resizebox{\hsize}{!}{\includegraphics{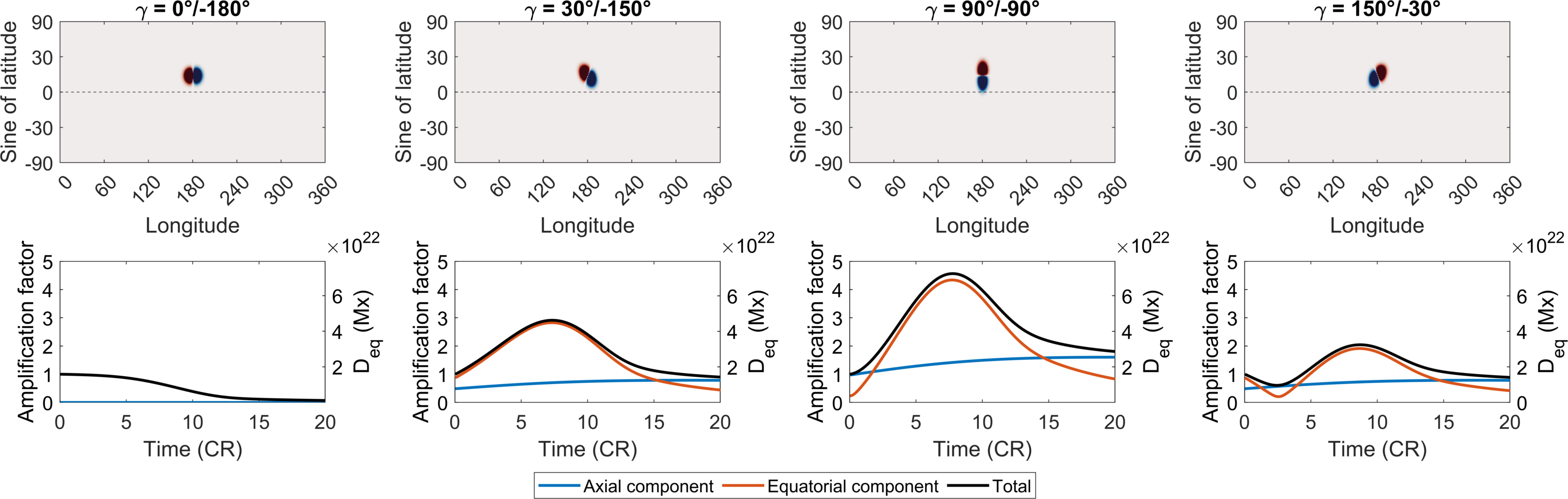}}
\caption{Effect of tilt angle on BMR dipole components. Top row show shows the BMR and the bottom row the evolution of dipole components. The black line shows the evolution of the total dipole strength while the blue and orange lines show the evolution of axial and equatorial components.
}\label{fig:BMRExamples}
\end{figure*}

To highlight the role of tilt angle in producing the different evolution patterns shown in Fig.~\ref{fig:ARExamples}, Fig.~\ref{fig:BMRExamples} shows the evolution of the dipole components for a synthetic BMR with four initial tilt angles: 0\degree{}, 30\degree{}, 90\degree{}, and 150\degree{}.
The other BMR parameters were set to $\Phi=1.66\times10^{23}$~Mx, $\lambda=13.5\degree{}$, and $\rho=11\degree{}$, corresponding to the properties of HMI SHARP 4698 (NOAA 12192) shown in the first column of Fig.~\ref{fig:ARExamples}.
We use the same y-axis scale in all plots to visually highlight the effect of tilt angle on the amplification of the active region dipole.

Figure~\ref{fig:BMRExamples} shows that, by changing the tilt angle, we can qualitatively reproduce the different evolution profiles shown in Fig.~\ref{fig:ARExamples}.
The dipole strength of a BMR with zero tilt angle decreases monotonically after emergence, unlike the other BMRs, whose equatorial dipole components are amplified considerably.
Because of the lack of amplification, the maximum dipole strength of the zero-tilt BMR is also much smaller than that of the regions with non-zero tilt.

The second and third columns of Fig.~\ref{fig:BMRExamples} show that increasing the tilt angle completely changes the dipole evolution.
Instead of decaying immediately, the equatorial dipole strengthens for about six rotations before starting to decay.
Although the initial equatorial component of the BMR with 90\degree{} tilt is almost zero (there is a small contribution from the latitude), its maximum strength is the largest of the four.
While the axial component is also non-zero, the amplification of the total dipole strength is driven almost entirely by the equatorial component.

The fourth column of Fig.~\ref{fig:BMRExamples} shows a BMR with anti-Joy tilt angle of -30\degree{} (or equivalently anti-Joy-anti-Hale tilt angle of 150\degree{}).
The evolution of this BMR differs from the others as the equatorial and total dipole strengths initially decrease, but then increase after reaching a local minimum.
Because of this initial decrease, the maximum dipole strength of this BMR is smaller than that of the BMR with 30\degree{} tilt angle in the second column.
Thus, the evolution of the equatorial dipole component of an anti-Joy BMR with -30\degree{} tilt differs from that of a regular BMR with 30\degree{} tilt while their axial evolution is identical.

\begin{figure} 
\resizebox{\hsize}{!}{\includegraphics{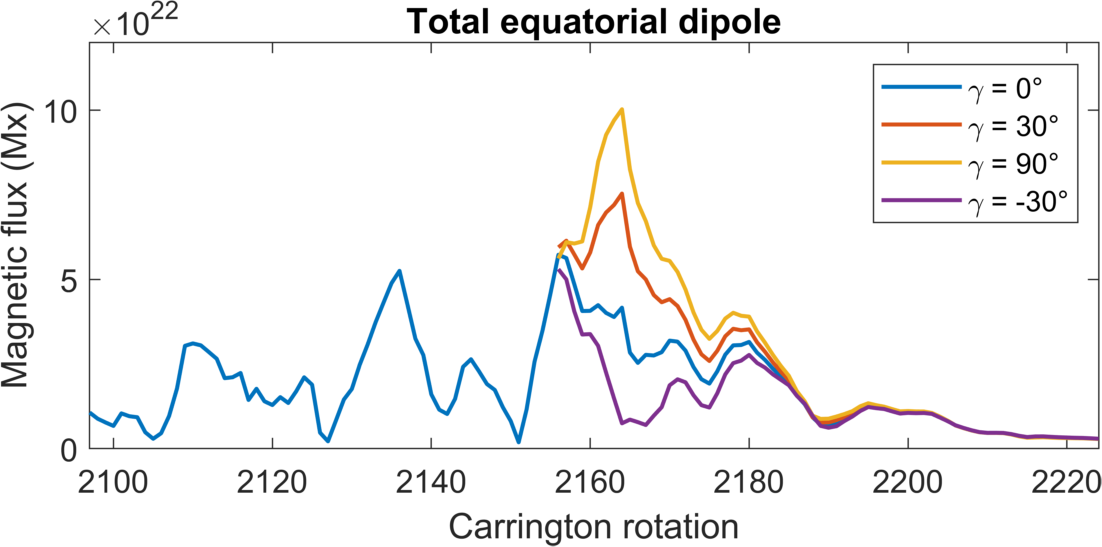}}
\caption{Effect of tilted NOAA 12192 on the net equatorial dipole. Blue line shows the simulation with original tilt, 0\degree{}, while the curves show the effect of different tilts.
}\label{fig:RotatedNOAA12192}
\end{figure}

Figure~\ref{fig:RotatedNOAA12192} demonstrates the effect that tilting NOAA 12192 has on the net equatorial component of the global solar magnetic field in a full simulation of solar cycle 24.
Tilting NOAA 12192 by 90\degree{} almost doubles the strength of the net equatorial dipole.
Smaller 30\degree{} tilt also introduces a considerable increase in the equatorial dipole strength with respect to the simulation with untilted NOAA 12192.
Contrastingly, NOAA 12192 with an anti-Joy angle of -30\degree{} causes a significant weakening that almost cancels the equatorial component.
The weakening of the net equatorial dipole happens because the differential rotation causes the faster rotating trailing polarity flux to overtake the slower rotating leading polarity flux, which flips the direction of the equatorial component by 180\degree{}, essentially turning the anti-Joy region into anti-Hale region.
\citet{Pal2023} also describe the transformation from an anti-Joy to an anti-Hale configuration (see their Figure~4), although in the context of the axial dipole moment.

\section{Effect of BMR parameters}\label{sec6:EffectOfParameters}
\subsection{Effect of tilt angle and latitude}
We studied the effect of BMR parameters by modeling the evolution of BMRs with DFT simulations over a range of parameter values.
In our simulations, the tilt angle runs from 0\degree{} to 180\degree{} in steps of 1\degree{}, the latitude from 0\degree{} to 70\degree{} in steps of 1\degree{}, and the separation from 5\degree{} to 15\degree{} in steps of 5\degree{}.
We kept the longitude fixed at 180\degree{} and magnetic flux fixed at $1.66\times10^{23}$~Mx.

\begin{figure*} 
\resizebox{\hsize}{!}{\includegraphics{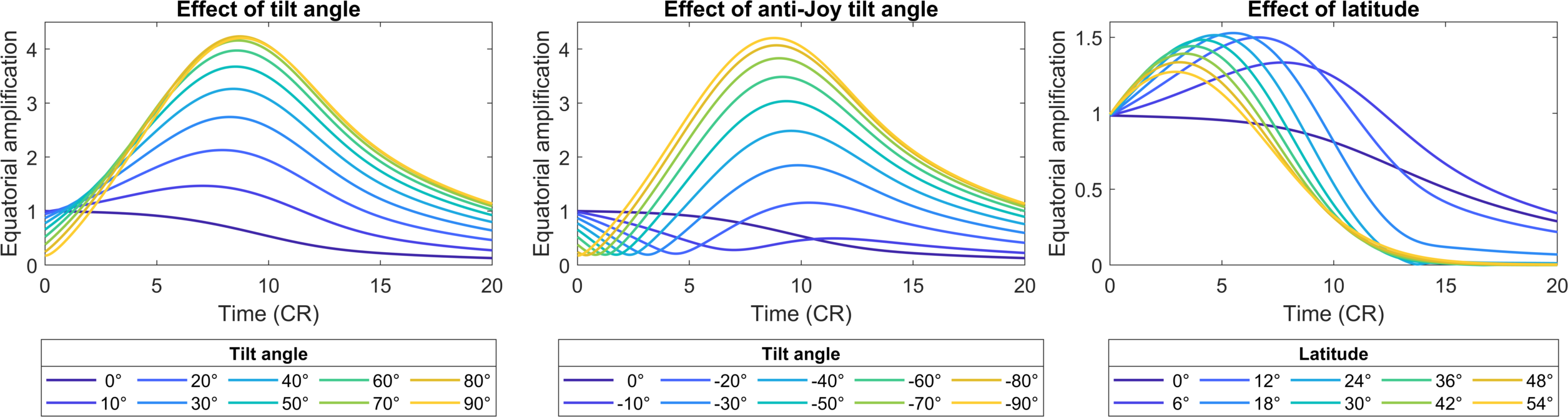}}
\caption{Effect of tilt angle and latitude on BMR dipole evolution. Left panel: effect of regular tilt angle. Middle panel: effect of anti-Joy tilt angle. Right panel: effect of latitude. The latitude is held constant at 10\degree{} in the left and middle panels, while the tilt angle is held constant at 10\degree{} in the right panel.  
}\label{fig:EffectOfTiltAndLatitude}
\end{figure*}

Figure~\ref{fig:EffectOfTiltAndLatitude}a shows the equatorial amplification factor as a function of time for BMRs with different tilt angles, with latitude held constant at $10$\degree{} and the separation fixed at 10\degree{}.
As already hinted by Fig.~\ref{fig:BMRExamples}, the tilt angle has a strong effect on the peak dipole strength, with larger tilt angles leading to stronger dipole amplification.
A tilt angle of 40\degree{} leads to an approximately three-fold increase in the peak dipole strength.
The maximum amplification factor for the largest tilt angles ($>70\degree{}$) is about 4.
The strength of the equatorial dipole increases from the time of emergence for all tilt angles except zero tilt, which leads to a monotonic decrease in the dipole strength.
The timing of the peak dipole strength depends only weakly on the tilt angle, being eight rotations for $\gamma=10\degree{}$ and nine rotations for larger tilt angles.

Figure~\ref{fig:EffectOfTiltAndLatitude}b shows the equatorial amplification factor as a function of time for anti-Joy BMRs with different tilt angles, with latitude held constant at 10\degree{} and the separation fixed at 10\degree{}.
Unlike regular Joy regions, the equatorial dipole of anti-Joy regions first weakens with time before a subsequent strengthening.
Only regions with anti-Joy tilt angles larger than 20\degree{} experience amplification relative to the initial dipole strength.
Larger tilt angles lead to an earlier turnover, after which the evolution is similar to that of regular active regions, with larger tilt angles again leading to a larger peak dipole strength.

Figure~\ref{fig:EffectOfTiltAndLatitude}c shows the equatorial amplification factor as a function of time for BMRs emerging at different latitudes, with the tilt angle held constant at 10\degree{} and the separation fixed at 10\degree{}.
While the latitude has some effect on the peak dipole strength, its effect is much smaller than that of the tilt angle.
The relation between latitude and peak dipole strength is also non-monotonic: the peak strength increases with increasing latitude at lower latitudes, but decreases above 15\degree{} latitude.
On the other hand, the latitude has a substantial effect on the timing of the peak dipole strength.
The time required for an active region dipole to reach its peak strength decreases with increasing latitude.
Active regions emerging at latitude 5\degree{} reach the peak dipole strength in 10 rotations, whereas active regions emerging above 30\degree{} reach the peak strength in only 4 rotations.
Active regions emerging at the equator do not follow this pattern, as their equatorial dipole decays monotonically similarly to active regions with zero tilt angle, although the decay happens more gradually.

\subsection{Amplification maps}
\begin{figure*} 
\resizebox{\hsize}{!}{\includegraphics{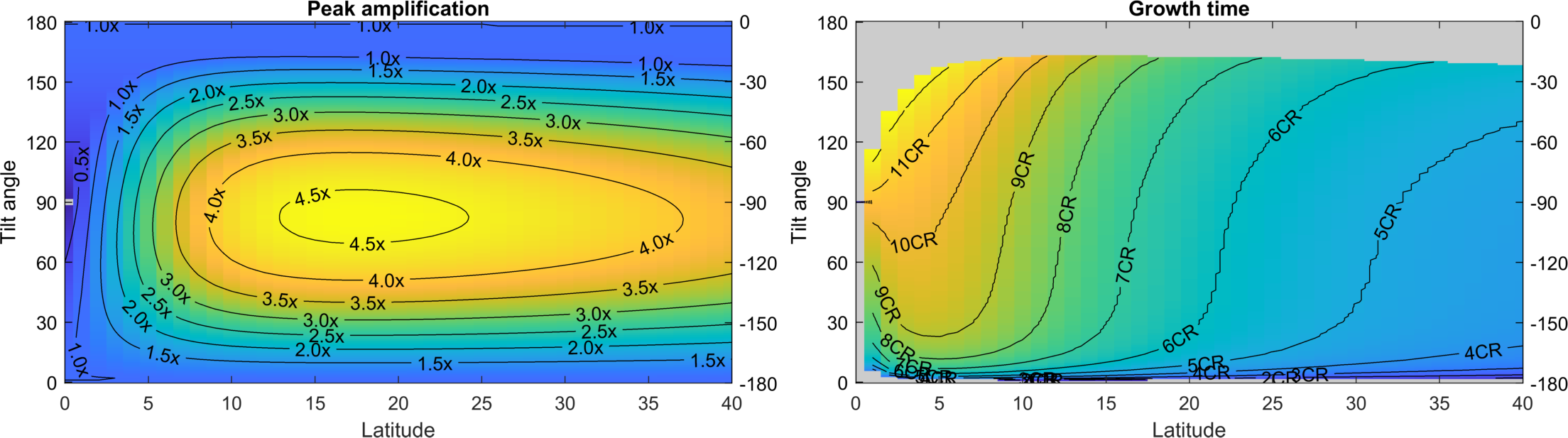}}
\caption{Effect of latitude and tilt angle on the maximum equatorial amplification. (a) Maximum equatorial amplification factor $A_{\rm max}$. (b) Growth time $t_{\rm growth}$. Gray region corresponds to parameter space where the equatorial dipole starts to immediately decay. Y-axis shows positive tilt angles on left and corresponding negative tilt angles on right.
}\label{fig:MaxAmplificationMaps}
\end{figure*}

Figure~\ref{fig:MaxAmplificationMaps}a shows the maximum amplification of the equatorial dipole for all simulated latitudes and tilt angles.
Tilt angles larger than 90\degree{} correspond to anti-Joy tilt angles.
The maximum amplification is about 4.6 times the initial total dipole strength, with the largest amplification occurring for regions with tilt angles close to 90\degree{}.
The maximum is not reached exactly at 90\degree{}, but at 82\degree{}.
Latitude has the largest effect on active regions with large tilt angles.
The greatest amplification is achieved for active regions emerging at latitudes around 15\degree{}--20\degree{}.

Figure~\ref{fig:MaxAmplificationMaps}b shows the time that it takes for the equatorial dipole to reach its maximum strength.
For a fixed tilt angle, higher latitude BMRs typically reach their maximum equatorial strength faster.
For a fixed latitude the timing of equatorial maximum increases slightly with increasing tilt angle.
Anti-Joy regions reach maximum strength later than regular Joy regions.

\begin{figure*} 
\resizebox{\hsize}{!}{\includegraphics{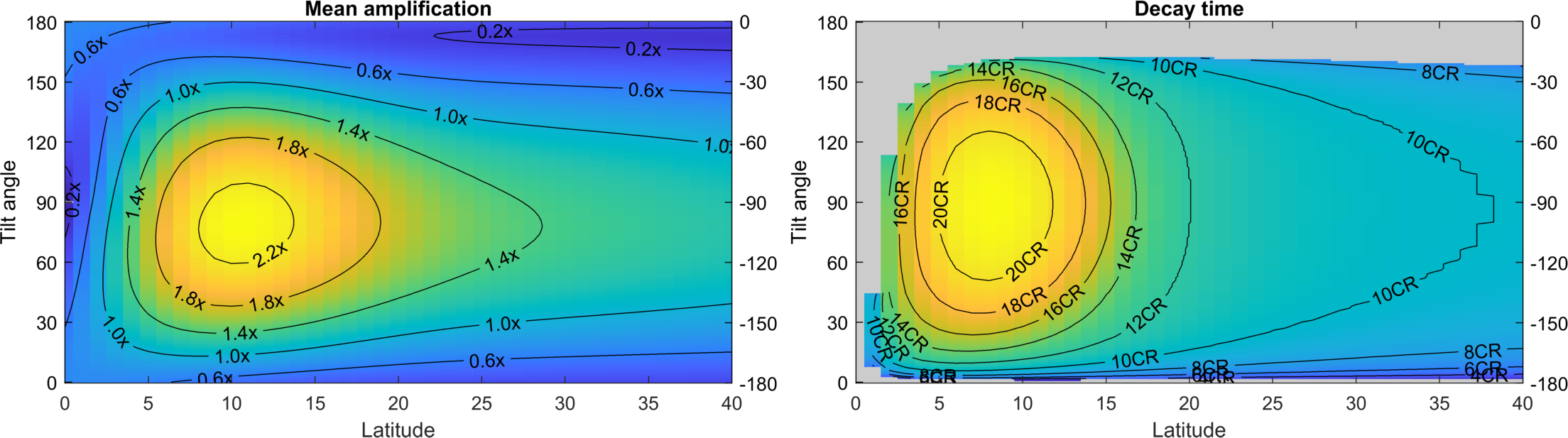}}
\caption{Effect of latitude and tilt angle on the mean equatorial amplification. (a) Mean amplification factor $\langle A \rangle $. (b) Decay time $t_{\rm decay}$. Gray region corresponds to parameter range where the strength of equatorial dipole is always smaller than the initial total dipole strength. Y-axis shows positive tilt angles on left and corresponding negative tilt angles on right.
}\label{fig:MeanAmplificationMaps}
\end{figure*}

In contrast to the peak amplification, Fig.~\ref{fig:MeanAmplificationMaps} shows the mean equatorial amplification factor during the first 22 rotations as a function of latitude and tilt angle emphasizing the overall strengthening of the equatorial dipole.
Similarly to the maximum strength, large tilt angles correspond to a large overall strengthening.
However, the range of latitudes with a large overall strengthening is significantly smaller than for the maximum amplification.
The largest strengthening is also achieved at lower latitudes than the largest maximum strength.

The reason for this different behavior can be seen from Fig.~\ref{fig:MeanAmplificationMaps}b, which shows the decay time, $t_{\rm decay}$, as a function of latitude and tilt angle.
The mean amplification is largest for regions with longest decay time.
While the equatorial component experiences large amplification also at high latitudes, the equatorial component of high-latitude regions decays much faster than low-latitude regions.

\subsection{Effect of polarity separation}
\begin{figure*} 
\resizebox{\hsize}{!}{\includegraphics{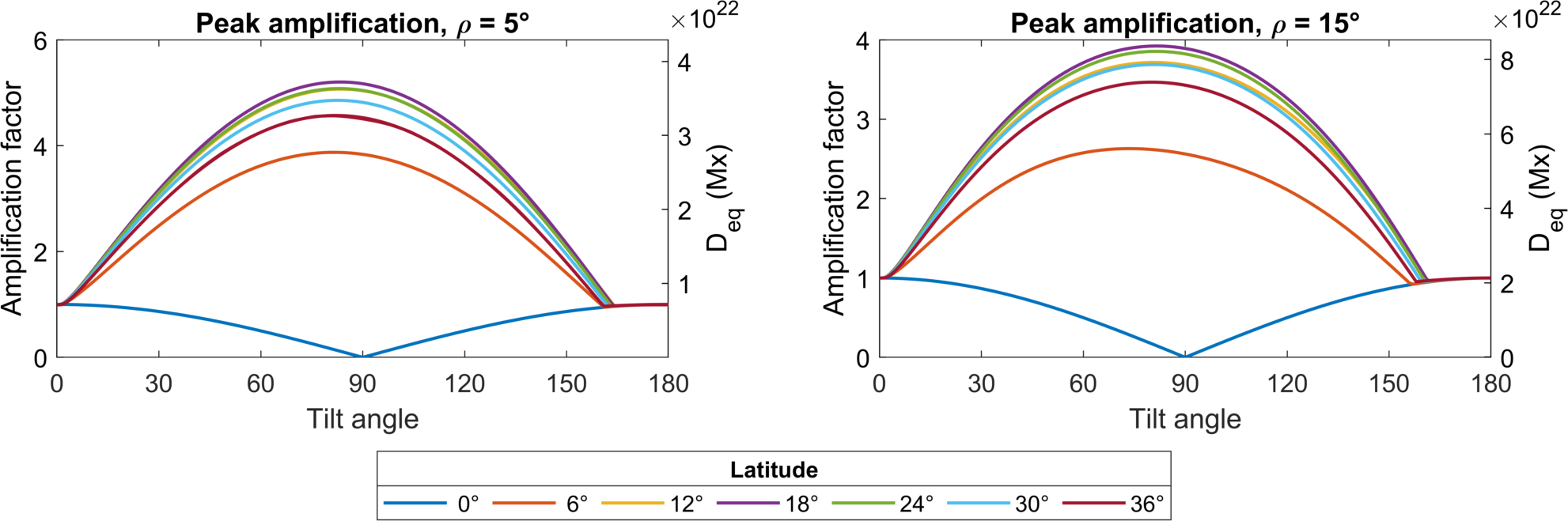}}
\caption{Maximum equatorial amplification as a function of tilt angle for different emergence latitudes. (a) 5\degree{} polarity separation. (b) 15\degree{} polarity separation. The left axis shows the amplification factor while the right-axis shows absolute strength.
}\label{fig:EffectOfSeparation}
\end{figure*}

Figure~\ref{fig:EffectOfSeparation} shows the maximum equatorial dipole strength as a function of tilt angle for different emergence latitudes for two different polarity separations, 5\degree{} and 15\degree{}.
Comparing panels a and b shows that active regions with smaller polarity separation experience larger amplification.
However, in terms of absolute strength, the equatorial component of regions with larger polarity separation grows to almost twice as large.
The reason is that initial dipole strength of regions with large polarity separation is larger than regions with smaller separation.
This reduces the amplification factor, but also reduces cancellation between the opposite polarities.
There is also a clear asymmetry between moderate regular tilt angles ($\gamma < 20\degree{}$) and moderate anti-Joy tilt angles ($\gamma > 160\degree{}$ or $\gamma > -20\degree{}$).
Active regions with moderate anti-Joy tilt angles do not experience amplification of the equatorial dipole.
This occurs because for moderate anti-Joy tilt angles the opposite-polarity fluxes remain close to each other for longer as the trailing-polarity flux overtakes the leading-polarity flux, allowing diffusion to annihilate much of the active region flux.

\section{Active regions of solar cycles 24 and 25}\label{sec7:SHARPS}
\begin{figure*} 
\resizebox{\hsize}{!}{\includegraphics{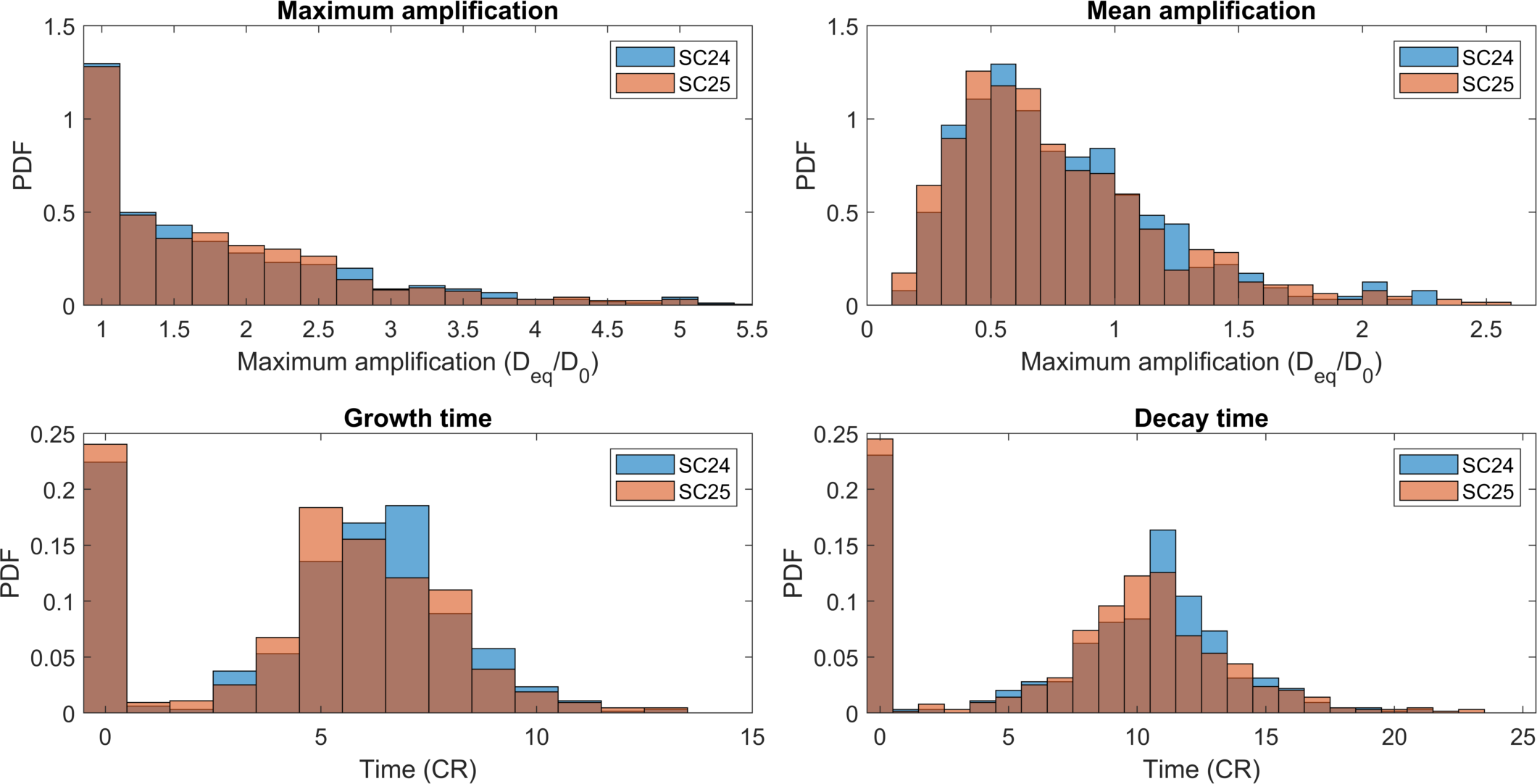}}
\caption{Histograms of equatorial dipole amplification measures and evolution times for HMI SHARPs of solar cycles 24 and 25. The four panels show the distributions of maximum amplification, mean amplification, time of maximum, and decay time.
}\label{fig:SHARPHistograms}
\end{figure*}

Figure~\ref{fig:SHARPHistograms} shows the distributions of the quantities studied in the previous section for HMI SHARPs of solar cycles (SCs) 24 and 25.
Overall, the results are similar for the two cycles.
The median and 75\% quantile of the maximum amplification are 1.43 and 2.19 for  SC24 and 1.41 and 2.18 for SC25.
The corresponding values for the mean amplification are 0.73 and 1.03 for SC24 and 0.72 and 1.03 for SC25.
About one fourth of active regions do not experience any amplification of the equatorial dipole component.
For those that do, the equatorial dipole typically reaches its maximum strength after about 5--7 rotations and decays after about 13--14 rotations.

There is some indication that active regions in SC25 evolve slightly faster than those in SC24, in the sense that both the rise time and the decay time are somewhat shorter.
This difference also remains when we restrict the comparison to active regions from the same phase of the solar cycle.
However, this effect is rather weak.
The mean growth times are $175\pm3$ days for SC24 and $170\pm4$  days for SC25, while the corresponding decay times are $293\pm5$ days and $290\pm6$ days, where the uncertainties denote bootstrap estimates of the two-sigma limits

\begin{figure*} 
\resizebox{\hsize}{!}{\includegraphics{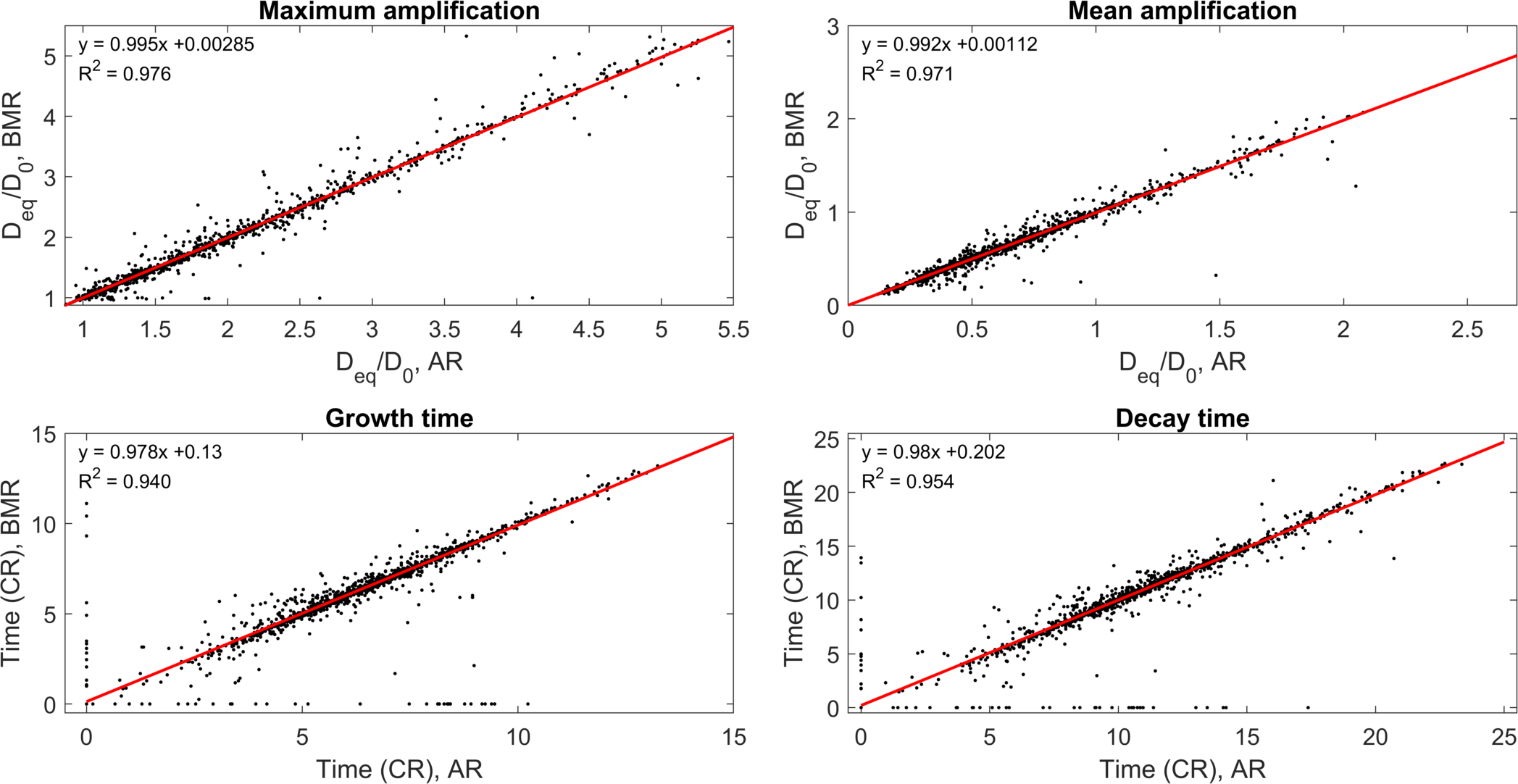}}
\caption{Comparison of equatorial amplification measures derived from original active regions and from their BMR approximations. The four panels show scatter plots of the maximum amplification, mean amplification, time of maximum, and decay time of the BMRs versus the corresponding active regions. The red line shows the best-fitting least-squares linear relation. In the lower row, points with zero growth time or zero decay time were excluded from the fit.
}\label{fig:ARvsBMR}
\end{figure*}

Figure~\ref{fig:ARvsBMR} compares the amplification measures derived from the HMI SHARPs and from corresponding BMRs.
Overall, the BMR approximation agrees well with the original SHARPs for all the studied quantities.
There are some active regions for which the growth or decay time is zero for the BMR but non-zero for the original SHARP, and vice versa.
Although the disagreement in the time of maximum or decay time can be large for such regions, most of them do not stand out as outliers in the maximum or mean amplification plots.
This result indicates a difference between BMR and (effective) AR tilt angle, where where one representation has zero tilt while the other has a small but non-zero tilt.
As shown in Figs.~\ref{fig:MaxAmplificationMaps} and \ref{fig:MeanAmplificationMaps}, there is a sharp transition from zero growth time to much longer time when tilt angle only slightly deviates from zero, while the mean and maximum amplification change only slightly.
Such differences can lead to large differences in the growth and decay times, even though the differences in dipole strength itself remains small.

\begin{figure*} 
\resizebox{\hsize}{!}{\includegraphics{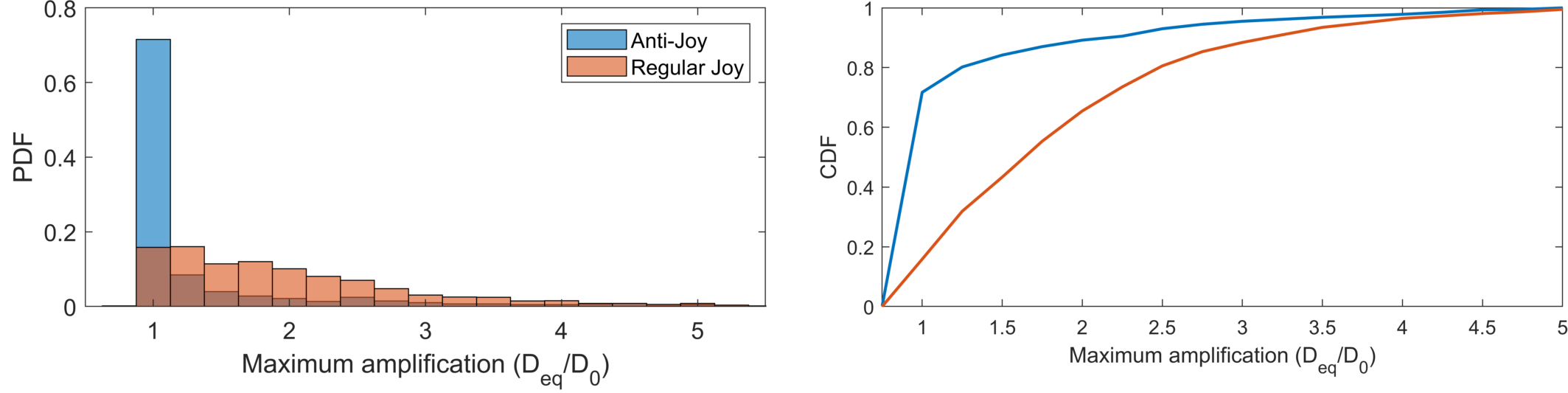}}
\caption{Histograms of the maximum equatorial amplification factor for regular Joy (blue) and anti-Joy (orange) active regions. (a) Probability density function. (b) Cumulative distribution function.
}\label{fig:AJHistogram}
\end{figure*}

Figure~\ref{fig:AJHistogram} compares the maximum amplification of regular Joy ($0\degree{} <\gamma < 90\degree{}$) and anti-Joy ($-90\degree{} <\gamma < 0\degree{}$) active regions.
There is a significant difference between two types of regions.
Most, 69\%, of the anti-Joy regions do not experience any amplification of the equatorial dipole component, while the same is true for only 4\% of the regular Joy regions.
Evidently, anti-Joy regions are much less likely than regular Joy regions to produce strong equatorial dipole amplification.
This result is in agreement with the results of synthetic BMR simulations from previous Sect.~\ref{sec6:EffectOfParameters}, which show that small to moderate anti-Joy tilt angles do not produce equatorial amplification.

\section{Discussion}\label{sec:Discussion}
We have studied the effect of tilt angle, latitude, and the polarity separation on the evolution of equatorial dipole component of BMRs. 
We simulated the dipole evolution with the DFT model \citep{Tahtinen2026b}, a matrix formulation of the SFT model, which produces equivalent results while being orders of magnitude faster.
We quantified the dipole evolution by studying max and mean amplification as well as growth and decay time of the equatorial dipole component.
We also studied the equatorial dipole evolution of active regions from the synoptic HMI SHARP database \citep{Yeates2020} and their BMR approximation.

\subsection{The role of BMR parameters}
The differential rotation drives the amplification of the equatorial dipole, while the meridional flow and diffusion both act to weaken it.
For the amplification mechanism to work a non-zero tilt angle is crucial, since there must be latitudinal separation between the opposite polarities.
Larger tilt angles lead to larger latitudinal separation by allowing more flux to be sheared apart, which increases the peak dipole strength.
The maximum amplification does not correspond to exactly 90\degree{} tilt angle but somewhat smaller tilt angle around 80\degree{}.
This is likely due to interplay between the diffusion and differential rotation.
The polarities of a BMR with 90\degree{} tilt angle spend a somewhat longer time in the vicinity of each other, which allows the diffusion to annihilate more flux.
However, these differences for largest tilt angles ($>70\degree{}$) are marginal, as can be seen from Fig.~\ref{fig:EffectOfTiltAndLatitude}a.

Figures~\ref{fig:MaxAmplificationMaps}b and \ref{fig:MeanAmplificationMaps}b show that while the latitude has a weaker effect on the peak dipole strength, it controls the timescale at which the equatorial dipole grows and decays.
At low latitudes the equatorial component evolves more slowly, whereas at high latitudes it grows and decays rapidly.
In principle the amplification is more effective at higher latitudes where the differential shear is larger and opposite polarities are more effectively sheared apart.
However, because the physical longitudinal distance decreases also at higher latitudes, the optimal latitude at which these effects balance out is around $\lambda=\pm20\degree{}$ latitude (see Figs.~\ref{fig:EffectOfTiltAndLatitude}c and \ref{fig:MaxAmplificationMaps}a).

Because the faster amplification is related to faster decay of the equatorial dipole, the latitudes associated with strongest amplification differ from those of producing strongest overall amplification over the lifetime of the equatorial dipole.
The largest overall strengthening is produced by somewhat lower latitudes around 10\degree{}.
At lower latitudes the latitudinal shear is weaker and it takes longer time for the magnetic flux to wrap around the Sun and to be transported to the polar regions where equatorial dipole is quickly annihilated by diffusion \citep{WangSheeley1991}.

The fact that the equatorial dipole of low-latitude regions persists for a longer time might have implications for the evolution of the net dipole and global solar magnetic field.
Since active regions typically emerge at lower-latitudes during the declining phase of the solar cycle, the increased lifetime of equatorial dipole might be relevant for the formation of long-lived low-latitude coronal holes, commonly observed during the declining phase of the cycle.

We find that larger polarity separation increases the equatorial dipole in absolute sense while decreasing the amplification factor.
The absolute strength increases, because larger polarity separation decreases diffusion between BMRs poles.
On the other hand, since large polarity separation also increases the initial dipole strength, BMRs with large separation experience relatively smaller amplification of the equatorial component.

\subsection{Anti-Joy regions}
Figure~\ref{fig:EffectOfTiltAndLatitude} shows that anti-Joy regions display a peculiar evolution where the equatorial dipole initially weakens, then strengthens and finally decays away.
The reason is that anti-Joy regions emerge with their leading polarity on higher latitudes than trailing polarity.
Because the differential rotation is faster on lower latitudes, the trailing polarity catches up to the leading polarity weakening the strength of the equatorial dipole.
Eventually, the trailing polarity overtakes the leading polarity, after which the equatorial dipole starts to grow as the latitudinal shear starts to work for the equatorial component instead of against it.
In this process, the equatorial component flips its direction by 180\degree{}, essentially turning the anti-Joy region into an anti-Hale region.
In a similar sense, anti-Joy-anti-Hale regions become regular regions.
This transformation occurs faster for higher-latitude regions because of the stronger latitudinal shear, resulting in an earlier reversal of the equatorial dipole.

The effect of this longitudinal polarity reversal is also seen in Fig.~\ref{fig:RotatedNOAA12192}.
Tilting NOAA 12192 30\degree{} anti-Joy -wise significantly decreases the strength of the global net equatorial dipole for over a year.
The reason is that the direction of the global dipole was closely aligned with the equatorial dipole of NOAA 12192 at the time of its emergence \citep{WangSheeley2015,Tahtinen2026a}.
Since anti-Joy tilt angle effectively flips the direction of the equatorial dipole, the magnetic flux of tilted NOAA 12192 acts to cancel the net equatorial dipole although the two are initially aligned.

Important consequence of this longitudinal polarity reversal is that the equatorial dipole of a BMR with an anti-Joy angle is not amplified as effectively as of regularly tilted BMR. 
This can be best seen by comparing the opposite ends of the tilt angle range in Fig.~\ref{fig:EffectOfSeparation}.
On the right hand side of both panels, there is a flat region corresponding to BMRs with small anti-Joy tilt angles that experience no amplification.
This happens because diffusion can effectively annihilate much of the active region flux during the time that the lower latitude flux overtakes the leading flux.
Since the tilt angles of the observed BMRs are typically rather small, this asymmetry between regular and anti-Joy regions has large effect on their amplification distribution. 
Figure~\ref{fig:AJHistogram} shows that most (69\%) of the anti-Joy regions do not experience amplification of equatorial dipole, whereas the same is true only for 4\% of regular tilt regions.

\subsection{AR vs BMR}
Figure~\ref{fig:SHARPHistograms} shows that the evolution statistics of HMI SHARPs agree well with the statistics calculated from the corresponding BMRs.
This indicates that, at least for metrics studied here, the BMR approximation is enough to capture the evolution of the equatorial dipole.
This is important for forecasting purposes as well as for reconstructions of historical magnetic fields from sunspot records as they do not contain enough information about the exact structure of active regions.

The biggest differences between SHARPs and BMRs are seen in growth and decay times as there are some regions, for which only the other experiences equatorial amplification while the other does not.
The number of such regions is larger than number of significant outliers in the two upper panels, indicating that despite large differences in timescales, the overall evolution of the regions does not necessarily differ that much.
Figures \ref{fig:MaxAmplificationMaps}a and \ref{fig:MeanAmplificationMaps}b and histograms of Fig.~\ref{fig:SHARPHistograms} show, the transition from zero decay time to larger ones is not continuous but jumps from zero to almost a year.
This means that for regions near this parameter boundary, small changes in tilt angle or latitude can cause a large differences in peak and decay times while the difference in the amplitude might still be rather small.

\subsection{Connection to solar and geomagnetic activity}
Figure~\ref{fig:ARvsBMR} shows that the differences in evolution statistics between SC24 and SC25 appear modest.
There is some indication that active regions in SC25 evolve slightly faster than those in SC24, but the effect is weak.
Although the observed differences between SC24 and SC25 are statistically weak, such differences might still arise between cycles of different activity levels.
In particular, tilt and latitude quenching, which are proposed regulatory mechanisms for solar dynamo, relate the BMR tilt angles and latitudes to amplitude of the solar cycle \citep{Cameron2010,Jiang2020,Petrovay2020,Talafa2022}.

Tilt quenching refers to the anti-correlation between active region tilt angles and cycle amplitude
\citep{Dasi-Espuig2010,McClintock2013,Jha2020,Jiao2021},
whereas latitude quenching refers to the positive correlation between active region emergence latitudes and cycle amplitude
\citep{Jiang2020,Talafa2022,Yeates2025}.
According to our results, tilt and latitude quenching could in principle increase the strength and lifetime of the equatorial dipole relative to cycle amplitude during weak cycles.

As the equatorial dipole is the main driver of the OSF and interplanetary magnetic field on intracyclic timescales \citep{WangSheeley2000a,WangSheeley2000b,WangSheeley2003,WangSheeley2006,Yoshida2023,Tahtinen2026a}, our results also indicate possible consequences for the relationship between solar and geomagnetic activity.
All else being equal, based on our results, we could expect the strength of the equatorial dipole and therefore OSF to increase with respect to cycle amplitude if active regions emerge at systematically lower latitudes or have systematically larger tilt angles.
Such systematic changes could also lead to increased geomagnetic activity relative to the solar activity, similar to what was observed in the late 20th century \citep{Lockwood1999,Lockwood2022a}.
The relative increase of geomagnetic activity indicates changes in the structure of the solar magnetic field that allowed the more efficient production of strong heliospheric magnetic field and fast solar wind streams, with smaller photospheric magnetic flux budget \citep{Stamper1999,Mursula2017}.
Because the global solar magnetic field is dominantly shaped by the active region magnetic flux, it is plausible that this divergence of solar and geomagnetic activity could be related to change in properties of active regions.

Interestingly, \citet{Yeates2025} find that the emergence latitudes of solar cycle 20, which was quite weak in terms of solar activity but rather strong in terms of geomagnetic activity, were abnormally low, especially at the southern hemisphere.
According to our results, the low emergence latitudes might have contributed to formation of large and persistent coronal holes observed in the declining phase of the solar cycle 20 \citep{Sheeley1978}.
This interpretation is supported by analysis of \citet{Sheeley1978}, who explicitly note these coronal holes persisted "...in large-scale magnetic regions that slowly evolved from low-latitude BMR's".

\section{Conclusions}\label{sec:Conclusions}
We have shown that the tilt angle and latitude of active regions have a large effect on the evolution of equatorial dipole.
The main effect of the tilt angle is to control the amount of equatorial amplification, by introducing a latitudinal flux separation that the latitudinal shear caused by differential rotation can then pull apart in longitude.
The latitude has only small effect on the amount of amplification, but plays a big part in controlling the timescale at which the equatorial dipole grows and decays.
Low-latitude active regions give birth for the longest living equatorial dipole, which might play a role in development of low-latitude coronal holes in the declining phase of the solar cycle.

We find that there is considerable asymmetry between regular and anti-Joy active regions.
The anti-Joy regions with moderate tilt angles ($<20\degree$) do not experience equatorial amplification, because the diffusion annihilates much of the active region flux, during the time that it takes for the trailing lower latitude flux to overtake the leading flux.
Dipole simulations of HMI SHARPs show that while 96\% of regular active regions experience amplification of the equatorial dipole, the same is true only for 31\% of anti-Joy regions.
This overtake by lower latitude flux also effectively turns anti-Joy regions into anti-Hale regions in terms of the direction of their equatorial dipole.

Finally, we suggest that systematic changes in emergence latitudes and tilt angles can lead to increase of geomagnetic activity relative to solar activity.
Our results indicate that unusually low emergence latitudes, could have been consequential for the formation of large and persistent coronal holes that led to relatively high geomagnetic activity in the declining phase of otherwise rather weak solar cycle 20.

\begin{acknowledgements}
Author acknowledges the financial support by the Research Council of Finland to the SOLEMIP (project no. 357249).
Author wishes to acknowledge CSC – IT Center for Science, Finland, for computational resources.
\end{acknowledgements}

\bibliography{bibliography} % your references Yourfile.bib

@ARTICLE{Bhowmik2018,
       author = {{Bhowmik}, Prantika and {Nandy}, Dibyendu},
        title = "{Prediction of the strength and timing of sunspot cycle 25 reveal decadal-scale space environmental conditions}",
      journal = {Nature Communications},
         year = 2018,
        month = dec,
       volume = {9},
          eid = {5209},
        pages = {5209},
          doi = {10.1038/s41467-018-07690-0},
archivePrefix = {arXiv},
       eprint = {1909.04537},
 primaryClass = {astro-ph.SR},
       adsurl = {https://ui.adsabs.harvard.edu/abs/2018NatCo...9.5209B}
}

@ARTICLE{Bobra2014,
       author = {{Bobra}, M.~G. and {Sun}, X. and {Hoeksema}, J.~T. and {Turmon}, M. and {Liu}, Y. and {Hayashi}, K. and {Barnes}, G. and {Leka}, K.~D.},
        title = "{The Helioseismic and Magnetic Imager (HMI) Vector Magnetic Field Pipeline: SHARPs - Space-Weather HMI Active Region Patches}",
      journal = {\solphys},
         year = 2014,
        month = sep,
       volume = {289},
       number = {9},
        pages = {3549-3578},
          doi = {10.1007/s11207-014-0529-3},
archivePrefix = {arXiv},
       eprint = {1404.1879},
 primaryClass = {astro-ph.SR},
       adsurl = {https://ui.adsabs.harvard.edu/abs/2014SoPh..289.3549B}
}

@ARTICLE{Cameron2010,
       author = {{Cameron}, R.~H. and {Jiang}, J. and {Schmitt}, D. and {Sch{\"u}ssler}, M.},
        title = "{Surface Flux Transport Modeling for Solar Cycles 15-21: Effects of Cycle-Dependent Tilt Angles of Sunspot Groups}",
      journal = {\apj},
         year = 2010,
        month = aug,
       volume = {719},
       number = {1},
        pages = {264-270},
          doi = {10.1088/0004-637X/719/1/264},
archivePrefix = {arXiv},
       eprint = {1006.3061},
 primaryClass = {astro-ph.SR},
       adsurl = {https://ui.adsabs.harvard.edu/abs/2010ApJ...719..264C}
}

@ARTICLE{Cameron2016,
       author = {{Cameron}, R.~H. and {Jiang}, J. and {Sch{\"u}ssler}, M.},
        title = "{Solar Cycle 25: Another Moderate Cycle?}",
      journal = {\apjl},
         year = 2016,
        month = jun,
       volume = {823},
       number = {2},
          eid = {L22},
        pages = {L22},
          doi = {10.3847/2041-8205/823/2/L22},
archivePrefix = {arXiv},
       eprint = {1604.05405},
 primaryClass = {astro-ph.SR},
       adsurl = {https://ui.adsabs.harvard.edu/abs/2016ApJ...823L..22C}
}

@ARTICLE{Cliver2011,
       author = {{Cliver}, E.~W. and {Ling}, A.~G.},
        title = "{The Floor in the Solar Wind Magnetic Field Revisited}",
      journal = {\solphys},
         year = 2011,
        month = dec,
       volume = {274},
       number = {1-2},
        pages = {285-301},
          doi = {10.1007/s11207-010-9657-6},
       adsurl = {https://ui.adsabs.harvard.edu/abs/2011SoPh..274..285C}
}

@ARTICLE{Dasi-Espuig2010,
       author = {{Dasi-Espuig}, M. and {Solanki}, S.~K. and {Krivova}, N.~A. and {Cameron}, R. and {Pe{\~n}uela}, T.},
        title = "{Sunspot group tilt angles and the strength of the solar cycle}",
      journal = {\aap},
         year = 2010,
        month = jul,
       volume = {518},
          eid = {A7},
        pages = {A7},
          doi = {10.1051/0004-6361/201014301},
archivePrefix = {arXiv},
       eprint = {1005.1774},
 primaryClass = {astro-ph.SR},
       adsurl = {https://ui.adsabs.harvard.edu/abs/2010A&A...518A...7D}
}

@ARTICLE{Ijima2017,
       author = {{Iijima}, H. and {Hotta}, H. and {Imada}, S. and {Kusano}, K. and {Shiota}, D.},
        title = "{Improvement of solar-cycle prediction: Plateau of solar axial dipole moment}",
      journal = {\aap},
         year = 2017,
        month = nov,
       volume = {607},
          eid = {L2},
        pages = {L2},
          doi = {10.1051/0004-6361/201731813},
archivePrefix = {arXiv},
       eprint = {1710.06528},
 primaryClass = {astro-ph.SR},
       adsurl = {https://ui.adsabs.harvard.edu/abs/2017A&A...607L...2I}
}

@ARTICLE{Jiang2014,
       author = {{Jiang}, J. and {Cameron}, R.~H. and {Sch{\"u}ssler}, M.},
        title = "{Effects of the Scatter in Sunspot Group Tilt Angles on the Large-scale Magnetic Field at the Solar Surface}",
      journal = {\apj},
         year = 2014,
        month = aug,
       volume = {791},
       number = {1},
          eid = {5},
        pages = {5},
          doi = {10.1088/0004-637X/791/1/5},
archivePrefix = {arXiv},
       eprint = {1406.5564},
 primaryClass = {astro-ph.SR},
       adsurl = {https://ui.adsabs.harvard.edu/abs/2014ApJ...791....5J}
}

@ARTICLE{Jiang2018,
       author = {{Jiang}, Jie and {Wang}, Jing-Xiu and {Jiao}, Qi-Rong and {Cao}, Jin-Bin},
        title = "{Predictability of the Solar Cycle Over One Cycle}",
      journal = {\apj},
         year = 2018,
        month = aug,
       volume = {863},
       number = {2},
          eid = {159},
        pages = {159},
          doi = {10.3847/1538-4357/aad197},
archivePrefix = {arXiv},
       eprint = {1807.01543},
 primaryClass = {astro-ph.SR},
       adsurl = {https://ui.adsabs.harvard.edu/abs/2018ApJ...863..159J}
}

@ARTICLE{Jiang2023,
       author = {{Jiang}, Jie and {Zhang}, Zebin and {Petrovay}, Krist{\'o}f},
        title = "{Comparison of physics-based prediction models of solar cycle 25}",
      journal = {Journal of Atmospheric and Solar-Terrestrial Physics},
         year = 2023,
        month = feb,
       volume = {243},
          eid = {106018},
        pages = {106018},
          doi = {10.1016/j.jastp.2023.106018},
archivePrefix = {arXiv},
       eprint = {2212.01158},
 primaryClass = {astro-ph.SR},
       adsurl = {https://ui.adsabs.harvard.edu/abs/2023JASTP.24306018J}
}

@ARTICLE{Pal2023,
       author = {{Pal}, Shaonwita and {Bhowmik}, Prantika and {Mahajan}, Sushant S. and {Nandy}, Dibyendu},
        title = "{Impact of Anomalous Active Regions on the Large-scale Magnetic Field of the Sun}",
      journal = {\apj},
         year = 2023,
        month = aug,
       volume = {953},
       number = {1},
          eid = {51},
        pages = {51},
          doi = {10.3847/1538-4357/acd77e},
archivePrefix = {arXiv},
       eprint = {2305.13145},
 primaryClass = {astro-ph.SR},
       adsurl = {https://ui.adsabs.harvard.edu/abs/2023ApJ...953...51P}
}

@ARTICLE{Petrovay2020,
       author = {{Petrovay}, Krist{\'o}f and {Nagy}, Melinda and {Yeates}, Anthony R.},
        title = "{Towards an algebraic method of solar cycle prediction. I. Calculating the ultimate dipole contributions of individual active regions}",
      journal = {Journal of Space Weather and Space Climate},
         year = 2020,
        month = aug,
       volume = {10},
          eid = {50},
        pages = {50},
          doi = {10.1051/swsc/2020050},
archivePrefix = {arXiv},
       eprint = {2009.02299},
 primaryClass = {astro-ph.SR},
       adsurl = {https://ui.adsabs.harvard.edu/abs/2020JSWSC..10...50P}
}

@ARTICLE{Schatten1978,
       author = {{Schatten}, K.~H. and {Scherrer}, P.~H. and {Svalgaard}, L. and {Wilcox}, J.~M.},
        title = "{Using Dynamo Theory to predict the sunspot number during Solar Cycle 21}",
      journal = {\grl},
         year = 1978,
        month = may,
       volume = {5},
       number = {5},
        pages = {411-414},
          doi = {10.1029/GL005i005p00411},
       adsurl = {https://ui.adsabs.harvard.edu/abs/1978GeoRL...5..411S}
}

@ARTICLE{Snodgrass1990,
       author = {{Snodgrass}, Herschel B. and {Ulrich}, Roger K.},
        title = "{Rotation of Doppler Features in the Solar Photosphere}",
      journal = {\apj},
         year = 1990,
        month = mar,
       volume = {351},
        pages = {309},
          doi = {10.1086/168467},
       adsurl = {https://ui.adsabs.harvard.edu/abs/1990ApJ...351..309S}
}

@ARTICLE{Svalgaard2005,
       author = {{Svalgaard}, Leif and {Cliver}, Edward W. and {Kamide}, Yohsuke},
        title = "{Sunspot cycle 24: Smallest cycle in 100 years?}",
      journal = {\grl},
         year = 2005,
        month = jan,
       volume = {32},
       number = {1},
          eid = {L01104},
        pages = {L01104},
          doi = {10.1029/2004GL021664},
       adsurl = {https://ui.adsabs.harvard.edu/abs/2005GeoRL..32.1104S}
}

@ARTICLE{Tahtinen2024,
       author = {{T{\"a}htinen}, Ismo and {Asikainen}, Timo and {Mursula}, Kalevi},
        title = "{Straight outta photosphere: Open solar flux without coronal modeling}",
      journal = {\aap},
         year = 2024,
        month = aug,
       volume = {688},
          eid = {L32},
        pages = {L32},
          doi = {10.1051/0004-6361/202451267},
archivePrefix = {arXiv},
       eprint = {2408.11525},
 primaryClass = {astro-ph.SR},
       adsurl = {https://ui.adsabs.harvard.edu/abs/2024A&A...688L..32T}
}

@ARTICLE{Upton2014,
       author = {{Upton}, Lisa and {Hathaway}, David H.},
        title = "{Predicting the Sun's Polar Magnetic Fields with a Surface Flux Transport Model}",
      journal = {\apj},
         year = 2014,
        month = jan,
       volume = {780},
       number = {1},
          eid = {5},
        pages = {5},
          doi = {10.1088/0004-637X/780/1/5},
archivePrefix = {arXiv},
       eprint = {1311.0844},
 primaryClass = {astro-ph.SR},
       adsurl = {https://ui.adsabs.harvard.edu/abs/2014ApJ...780....5U}
}

@ARTICLE{Upton2018,
       author = {{Upton}, Lisa A. and {Hathaway}, David H.},
        title = "{An Updated Solar Cycle 25 Prediction With AFT: The Modern Minimum}",
      journal = {\grl},
         year = 2018,
        month = aug,
       volume = {45},
       number = {16},
        pages = {8091-8095},
          doi = {10.1029/2018GL078387},
archivePrefix = {arXiv},
       eprint = {1808.04868},
 primaryClass = {astro-ph.SR},
       adsurl = {https://ui.adsabs.harvard.edu/abs/2018GeoRL..45.8091U}
}

@ARTICLE{WangSheeley1991,
       author = {{Wang}, Y. -M. and {Sheeley}, Jr., N.~R.},
        title = "{Magnetic Flux Transport and the Sun's Dipole Moment: New Twists to the Babcock-Leighton Model}",
      journal = {\apj},
         year = 1991,
        month = jul,
       volume = {375},
        pages = {761},
          doi = {10.1086/170240},
       adsurl = {https://ui.adsabs.harvard.edu/abs/1991ApJ...375..761W}
}

@ARTICLE{WangSheeley2000a,
       author = {{Wang}, Y. -M. and {Lean}, J. and {Sheeley}, Jr., N.~R.},
        title = "{The long-term variation of the Sun's open magnetic flux}",
      journal = {\grl},
         year = 2000,
        month = feb,
       volume = {27},
       number = {4},
        pages = {505-508},
          doi = {10.1029/1999GL010744},
       adsurl = {https://ui.adsabs.harvard.edu/abs/2000GeoRL..27..505W}
}

@ARTICLE{WangSheeley2000b,
       author = {{Wang}, Y. -M. and {Sheeley}, Jr., N.~R. and {Lean}, J.},
        title = "{Understanding the evolution of the Sun's open magnetic flux}",
      journal = {\grl},
         year = 2000,
        month = mar,
       volume = {27},
       number = {5},
        pages = {621-624},
          doi = {10.1029/1999GL010759},
       adsurl = {https://ui.adsabs.harvard.edu/abs/2000GeoRL..27..621W}
}

@ARTICLE{WangSheeley2002,
       author = {{Wang}, Y.-M. and {Sheeley}, N.~R.},
        title = "{Sunspot activity and the long-term variation of the Sun's open magnetic flux}",
      journal = {Journal of Geophysical Research (Space Physics)},
         year = 2002,
        month = oct,
       volume = {107},
       number = {A10},
          eid = {1302},
        pages = {1302},
          doi = {10.1029/2001JA000500},
       adsurl = {https://ui.adsabs.harvard.edu/abs/2002JGRA..107.1302W}
}

@ARTICLE{WangSheeley2003,
       author = {{Wang}, Y. -M. and {Sheeley}, Jr., N.~R.},
        title = "{On the Fluctuating Component of the Sun's Large-Scale Magnetic Field}",
      journal = {\apj},
         year = 2003,
        month = jun,
       volume = {590},
       number = {2},
        pages = {1111-1120},
          doi = {10.1086/375026},
       adsurl = {https://ui.adsabs.harvard.edu/abs/2003ApJ...590.1111W}
}

@ARTICLE{WangSheeley2006,
       author = {{Wang}, Y.-M. and {Sheeley}, Jr., N.~R. and {Rouillard}, A.~P.},
        title = "{Role of the Sun's Nonaxisymmetric Open Flux in Cosmic-Ray Modulation}",
      journal = {\apj},
         year = 2006,
        month = jun,
       volume = {644},
       number = {1},
        pages = {638-645},
          doi = {10.1086/503523},
       adsurl = {https://ui.adsabs.harvard.edu/abs/2006ApJ...644..638W}
}

@ARTICLE{WangSheeley2015,
       author = {{Sheeley}, Jr., N.~R. and {Wang}, Y. -M.},
        title = "{The Recent Rejuvenation of the Sun's Large-scale Magnetic Field: A Clue for Understanding Past and Future Sunspot Cycles}",
      journal = {\apj},
         year = 2015,
        month = aug,
       volume = {809},
       number = {2},
          eid = {113},
        pages = {113},
          doi = {10.1088/0004-637X/809/2/113},
       adsurl = {https://ui.adsabs.harvard.edu/abs/2015ApJ...809..113S}
}

@ARTICLE{Whitbread2018,
       author = {{Whitbread}, T. and {Yeates}, A.~R. and {Mu{\~n}oz-Jaramillo}, A.},
        title = "{How Many Active Regions Are Necessary to Predict the Solar Dipole Moment?}",
      journal = {\apj},
         year = 2018,
        month = aug,
       volume = {863},
       number = {2},
          eid = {116},
        pages = {116},
          doi = {10.3847/1538-4357/aad17e},
archivePrefix = {arXiv},
       eprint = {1807.01617},
 primaryClass = {astro-ph.SR},
       adsurl = {https://ui.adsabs.harvard.edu/abs/2018ApJ...863..116W}
}

@ARTICLE{Yeates2020,
       author = {{Yeates}, Anthony R.},
        title = "{How Good Is the Bipolar Approximation of Active Regions for Surface Flux Transport?}",
      journal = {\solphys},
         year = 2020,
        month = sep,
       volume = {295},
       number = {9},
          eid = {119},
        pages = {119},
          doi = {10.1007/s11207-020-01688-y},
archivePrefix = {arXiv},
       eprint = {2008.03203},
 primaryClass = {astro-ph.SR},
       adsurl = {https://ui.adsabs.harvard.edu/abs/2020SoPh..295..119Y}
}

@ARTICLE{Yeates2023,
       author = {{Yeates}, Anthony R. and {Cheung}, Mark C.~M. and {Jiang}, Jie and {Petrovay}, Kristof and {Wang}, Yi-Ming},
        title = "{Surface Flux Transport on the Sun}",
      journal = {\ssr},
         year = 2023,
        month = jun,
       volume = {219},
       number = {4},
          eid = {31},
        pages = {31},
          doi = {10.1007/s11214-023-00978-8},
archivePrefix = {arXiv},
       eprint = {2303.01209},
 primaryClass = {astro-ph.SR},
       adsurl = {https://ui.adsabs.harvard.edu/abs/2023SSRv..219...31Y}
}

@ARTICLE{Yeates2025,
       author = {{Yeates}, Anthony R. and {Bertello}, Luca and {Pevtsov}, Alexander A. and {Pevtsov}, Alexei A.},
        title = "{Latitude Quenching Nonlinearity in the Solar Dynamo}",
      journal = {\apj},
         year = 2025,
        month = jan,
       volume = {978},
       number = {2},
          eid = {147},
        pages = {147},
          doi = {10.3847/1538-4357/ad99d0},
archivePrefix = {arXiv},
       eprint = {2412.02312},
 primaryClass = {astro-ph.SR},
       adsurl = {https://ui.adsabs.harvard.edu/abs/2025ApJ...978..147Y}
}

@ARTICLE{Tahtinen2026a,
       author = {{T{\"a}htinen}, Ismo and {Asikainen}, Timo and {Mursula}, Kalevi},
        title = "{Active regions and the large-scale magnetic field of solar cycle 24}",
      journal = {\aap},
         year = 2026,
        month = feb,
       volume = {706},
          eid = {A235},
        pages = {A235},
          doi = {10.1051/0004-6361/202557466},
       adsurl = {https://ui.adsabs.harvard.edu/abs/2026A&A...706A.235T}
}

@ARTICLE{Tahtinen2026b,
       author = {{T{\"a}htinen}, Ismo and {Asikainen}, Timo and {Mursula}, Kalevi},
        title = "{Ultra-fast simulations of the solar dipole and open flux}",
      journal = {\aap},
         year = 2026,
        month = apr,
       volume = {708},
          eid = {L21},
        pages = {L21},
          doi = {10.1051/0004-6361/202659586},
archivePrefix = {arXiv},
       eprint = {2604.11342},
 primaryClass = {astro-ph.SR},
       adsurl = {https://ui.adsabs.harvard.edu/abs/2026A&A...708L..21T}
}

@ARTICLE{Yoshida2023,
       author = {{Yoshida}, Minami and {Shimizu}, Toshifumi and {Toriumi}, Shin},
        title = "{Which Component of Solar Magnetic Field Drives the Evolution of Interplanetary Magnetic Field over the Solar Cycle?}",
      journal = {\apj},
         year = 2023,
        month = jun,
       volume = {950},
       number = {2},
          eid = {156},
        pages = {156},
          doi = {10.3847/1538-4357/acd053},
archivePrefix = {arXiv},
       eprint = {2304.13347},
 primaryClass = {astro-ph.SR},
       adsurl = {https://ui.adsabs.harvard.edu/abs/2023ApJ...950..156Y}
}

@ARTICLE{Yoshida2026,
       author = {{Yoshida}, Minami and {Shimizu}, Toshifumi and {Toriumi}, Shin and {Iijima}, Haruhisa},
        title = "{Temporal Evolution of Sunspot Groups and Increase in the Open Flux during Solar Maximum in Cycle 24}",
      journal = {\apj},
         year = 2026,
        month = apr,
       volume = {1001},
       number = {1},
          eid = {23},
        pages = {23},
          doi = {10.3847/1538-4357/ae4c4c},
archivePrefix = {arXiv},
       eprint = {2602.24118},
 primaryClass = {astro-ph.SR},
       adsurl = {https://ui.adsabs.harvard.edu/abs/2026ApJ..1001...23Y}
}

@ARTICLE{Jiang2020,
       author = {{Jiang}, Jie},
        title = "{Nonlinear Mechanisms that Regulate the Solar Cycle Amplitude}",
      journal = {\apj},
         year = 2020,
        month = sep,
       volume = {900},
       number = {1},
          eid = {19},
        pages = {19},
          doi = {10.3847/1538-4357/abaa4b},
archivePrefix = {arXiv},
       eprint = {2007.07069},
 primaryClass = {astro-ph.SR},
       adsurl = {https://ui.adsabs.harvard.edu/abs/2020ApJ...900...19J}
}

@ARTICLE{Talafa2022,
       author = {{Talafha}, M. and {Nagy}, M. and {Lemerle}, A. and {Petrovay}, K.},
        title = "{Role of observable nonlinearities in solar cycle modulation}",
      journal = {\aap},
         year = 2022,
        month = apr,
       volume = {660},
          eid = {A92},
        pages = {A92},
          doi = {10.1051/0004-6361/202142572},
archivePrefix = {arXiv},
       eprint = {2112.14465},
 primaryClass = {astro-ph.SR},
       adsurl = {https://ui.adsabs.harvard.edu/abs/2022A&A...660A..92T}
}

@ARTICLE{McClintock2013,
       author = {{McClintock}, B.~H. and {Norton}, A.~A.},
        title = "{Recovering Joy's Law as a Function of Solar Cycle, Hemisphere, and Longitude}",
      journal = {\solphys},
         year = 2013,
        month = oct,
       volume = {287},
       number = {1-2},
        pages = {215-227},
          doi = {10.1007/s11207-013-0338-0},
archivePrefix = {arXiv},
       eprint = {1305.3205},
 primaryClass = {astro-ph.SR},
       adsurl = {https://ui.adsabs.harvard.edu/abs/2013SoPh..287..215M}
}

@ARTICLE{Jha2020,
       author = {{Jha}, Bibhuti Kumar and {Karak}, Bidya Binay and {Mandal}, Sudip and {Banerjee}, Dipankar},
        title = "{Magnetic Field Dependence of Bipolar Magnetic Region Tilts on the Sun: Indication of Tilt Quenching}",
      journal = {\apjl},
         year = 2020,
        month = jan,
       volume = {889},
       number = {1},
          eid = {L19},
        pages = {L19},
          doi = {10.3847/2041-8213/ab665c},
archivePrefix = {arXiv},
       eprint = {1912.13223},
 primaryClass = {astro-ph.SR},
       adsurl = {https://ui.adsabs.harvard.edu/abs/2020ApJ...889L..19J}
}

@ARTICLE{Jiao2021,
       author = {{Jiao}, Qirong and {Jiang}, Jie and {Wang}, Zi-Fan},
        title = "{Sunspot tilt angles revisited: Dependence on the solar cycle strength}",
      journal = {\aap},
         year = 2021,
        month = sep,
       volume = {653},
          eid = {A27},
        pages = {A27},
          doi = {10.1051/0004-6361/202141215},
archivePrefix = {arXiv},
       eprint = {2106.11615},
 primaryClass = {astro-ph.SR},
       adsurl = {https://ui.adsabs.harvard.edu/abs/2021A&A...653A..27J}
}

@ARTICLE{Stamper1999,
       author = {{Stamper}, R. and {Lockwood}, M. and {Wild}, M.~N. and {Clark}, T.~D.~G.},
        title = "{Solar causes of the long-term increase in geomagnetic activity}",
      journal = {\jgr},
         year = 1999,
        month = jan,
       volume = {104},
       number = {A12},
        pages = {28325-28342},
          doi = {10.1029/1999JA900311},
       adsurl = {https://ui.adsabs.harvard.edu/abs/1999JGR...10428325S}
}

@ARTICLE{Lockwood1999,
       author = {{Lockwood}, M. and {Stamper}, R. and {Wild}, M.~N.},
        title = "{A doubling of the Sun's coronal magnetic field during the past 100 years}",
      journal = {\nat},
         year = 1999,
        month = jun,
       volume = {399},
       number = {6735},
        pages = {437-439},
          doi = {10.1038/20867},
       adsurl = {https://ui.adsabs.harvard.edu/abs/1999Natur.399..437L}
}

@ARTICLE{Lockwood2022a,
       author = {{Lockwood}, Mike and {Owens}, Mathew J. and {Barnard}, Luke A. and {Scott}, Chris J. and {Frost}, Anna M. and {Yu}, Bingkun and {Chi}, Yutian},
        title = "{Application of historic datasets to understanding open solar flux and the 20th-century grand solar maximum. 1. Geomagnetic, ionospheric, and sunspot observations}",
      journal = {Frontiers in Astronomy and Space Sciences},
         year = 2022,
        month = sep,
       volume = {9},
          eid = {960775},
        pages = {960775},
          doi = {10.3389/fspas.2022.960775},
       adsurl = {https://ui.adsabs.harvard.edu/abs/2022FrASS...9.0775L}
}

@ARTICLE{Mursula2017,
       author = {{Mursula}, K. and {Holappa}, L. and {Lukianova}, R.},
        title = "{Seasonal solar wind speeds for the last 100 years: Unique coronal hole structures during the peak and demise of the Grand Modern Maximum}",
      journal = {\grl},
         year = 2017,
        month = jan,
       volume = {44},
       number = {1},
        pages = {30-36},
          doi = {10.1002/2016GL071573},
archivePrefix = {arXiv},
       eprint = {1612.04941},
 primaryClass = {physics.space-ph},
       adsurl = {https://ui.adsabs.harvard.edu/abs/2017GeoRL..44...30M}
}

@ARTICLE{Sheeley1978,
       author = {{Sheeley}, Jr., N.~R. and {Harvey}, J.~W.},
        title = "{Coronal holes, solar wind streams, and geomagnetic activity during the new sunspot cycle}",
      journal = {\solphys},
         year = 1978,
        month = sep,
       volume = {59},
       number = {1},
        pages = {159-173},
          doi = {10.1007/BF00154939},
       adsurl = {https://ui.adsabs.harvard.edu/abs/1978SoPh...59..159S}
}

\appendix
\section{Surface flux transport}\label{appendix}
We use a surface flux transport (SFT) model to model the evolution of the large-scale solar magnetic field (see, e.g., Yeates 2023 for review).
The SFT model describes the evolution of the radial magnetic field subject to differential rotation $\Omega(\theta)$, meridional flow $u_\theta(\theta)$, and turbulent diffusion $\eta$ with the induction equation
\begin{equation}
\frac{\partial B_r}{\partial t} + \nabla_h \cdot (\mathbf{u}_h B_r) = \eta \nabla_h^2 B_r + S,
\end{equation}
where $\mathbf{u}_h$ describes the horizontal flow velocity (due to $\Omega(\theta)$ and $u_\theta(\theta)$) and the magnetic flux emergence is modeled with the source term S.
The explicit form of the governing equation is
\begin{gather}
\begin{aligned}
    \frac{\partial B_r}{\partial t} = &-\frac{1}{R_{\odot} \sin \theta} \frac{\partial}{\partial \theta} \left( \sin \theta \, u_{\theta} B_r \right)
    - \Omega(\theta) \frac{\partial B_r}{\partial \phi} \\
    &+ \frac{\eta}{R_{\odot}^2 \sin \theta} \frac{\partial}{\partial \theta}  
    \left( \sin \theta \frac{\partial B_r}{\partial \theta} \right)
    + \frac{\eta}{R_{\odot}^2 \sin^2 \theta} \frac{\partial^2 B_r}{\partial \phi^2} + S,
\end{aligned}
\end{gather}
where we adopt the differential rotation profile of \citet{Snodgrass1990}
\begin{equation}
    \Omega(\theta) = 0.18 - 2.396 \cos^2(\theta) - 1.787 \cos^4(\theta) \quad [^{\circ} \text{day}^{-1}],
\end{equation}
and the meridional flow profile of \citet{Whitbread2018}

\begin{equation}
        u_{\theta}(\theta) = -u_0 \sin^p{\theta}\cos{\theta},
\end{equation}
with their shape parameter $p = 2.33$.
We set diffusivity to $\eta=350~\mathrm{km^2/s}$ and the meridional flow peak velocity to $u_0=11$~m/s, which \citep{Tahtinen2026a} found the best-fit values for the solar cycle 24, using the total vector sum as optimization metric.

\end{document}